\documentclass[twocolumn,showpacs,floats,superscriptaddress,10pt]{revtex4}
\usepackage{graphicx}
\usepackage{amsmath}
\usepackage{bm}

\begin{document}

\title{Charge pumping in monolayer graphene driven by a series of
time-periodic potentials}
\author{Zhenhua Wu}
\affiliation{SKLSM, Institute of Semiconductors, Chinese Academy of Sciences,\\
P.O. Box 912, 100083, Beijing, China}
\affiliation{CAE Team, Semiconductor R $\&$ D Center, Samsung Electronics Co., Ltd.
Gyeonggi-Do, Korea}
\email{zhwu@semi.ac.cn, Tel.: +82-010-7287-1816}

\author{J. Li}
\affiliation{Department of Physics, Semiconductor Photonics Research Center, Xiamen University, Xiamen 361005,
China}

\author{K. S. Chan}
\affiliation{Department of Physics and Materials Science, \\
City University of Hong Kong,\\
Tat Chee Avenue, Kowloon, Hong Kong, China}
\email{apkschan@cityu.edu.hk}

\begin{abstract}
We applied the Floquet scattering-matrix formalism to studying the
electronic transport properties in a mesoscopic Dirac system. Using the
method, we investigate theoretically quantum pumping driven by a series of
time-periodic potentials in graphene monolayer both in the adiabatic and
non-adiabatic regimes. Our numerical results demonstrate that adding
harmonic modulated potentials can break the time reversal symmetry when no
voltage bias is applied to the graphene monolayer. Thus, when the system is
pumped with proper dynamic parameters, these scatterers can produce a
nonzero dc pumped current. We also find that the transmission is
anisotropic as the incident angle is changed. \newline
\textbf{Keywords:} graphene, pumping, Floquet scattering-matrix
\end{abstract}

\pacs{72.10.-d,73.20.-r,73.23.-b,73.40.Gk}
\maketitle



\section{Introduction}

Quantum transport in periodically driven mesoscopic systems is currently of
great interest because of an increasing number of applications.~\cite%
{Switkes,Thouless,Buttiker,Niu,Wagner,Burmeister,Brouwer,Wang,Zhu,Torres,Torres2,Rivera}
Many research studies have been carried out on the mechanism of scattering
by time-dependent potentials in recent years. There exists a good
theoretical understanding of the tunnelling events in the presence of an
alternating (AC) field, i.e., the Floquet theory,~\cite%
{Shirley,Li,Moskalets,Moskalets2} which takes the photon-assisted transport
into account and allows us to convert the solution of a time-periodic Schr%
\"{o}dinger equation into a time-independent eigenvalue problem. For systems
with time-periodic potentials, inelastic tunnelling occurs when electrons at
appropriate incident Fermi energies undergo transitions between the central
band to several sidebands by means of photon emission or absorption,
referred to as photon-assisted tunneling (PAT). The central band just
corresponds to particles with the incident Fermi energy\ while the sidebands
correspond to particles which have gained or lost one or more modulation
energy quantum. Especially, if the potential is harmonic in time, the PAT
results in exchange of energy with electrons in units of modulation quantum $%
\hbar \omega $, with $\omega $ being the modulation frequency. An important
application of the PAT is quantum charge pump, which received considerable
theoretical investigation and had been observed experimentally.\cite%
{Switkes,Thouless,Niu,Wagner,Brouwer,Wang,Zhu,Torres,Moskalets,Moskalets2}
However most previous studies in this realm have been focused on the Schr%
\"{o}dinger system.

In Ref.\cite{Moskalets2}, the Floquet scattering theory was developed for
quantum pumping in mesoscopic semiconductors by Moskalets and B\"{u}ttiker.
This approach permits description of both adiabatic and non-adiabatic
regimes of pumping. It is the purpose of this work to applied the Floquet
scattering-matrix method to calculating quantum charge pumping in the Dirac
system. To this end, we extend the approach of Ref.\cite{Moskalets2} to
accommodate a novel two dimensional hexagonal carbon material, graphene.
Recently, the study of quantum pumps in Dirac system has been carried out in
one dimensional Luttinger liquid model~\cite{Agarwal} or weak pumps in the adiabatic regime~\cite{Prada,RZhu} and nonadiabatic regime~\cite{San-Jose,Torres3} in graphene  by using the Floquet scattering theory.
Graphene, a single layer of carbon atoms arranged in a
hexagonal lattice, exhibits abundant new physics and potential applications.~%
\cite{Novoselov2,Zhang,Novoselov1,Geim,Neto} Quantum transport properties in
graphene have attracted increasing attentions both for the fundamental
physics and potential applications in carbon-based nano-electronics devices.
The novel properties arise from the unique linear dispersion and the chiral
nature of its carriers near the Dirac points. At the $K$ ($K^{^{\prime }}$)
point of the Brillouin zone, the energy spectrum exhibits a linear
dispersion that can be well described by the massless Dirac equation with an
effective speed of $v_{F}\approx 10^{6}ms^{-1}$.~\cite{Neto} These
quasiparticles, called massless Dirac fermions, are quite different from the
standard electrons that we encounter in conventional two dimensional
electron gases in semiconductor heterostructures. Graphene monolayer is a
suitable testbed to examine the photon-assisted electron tunneling~\cite{Trauzette,Zeb}
and quantum pumps in the Dirac system. In addition, the high electron
mobility and long phase coherence present in graphene could in principle
enhance these interference effects (photon-assisted tunneling and quantum
pumps), thereby making possible its observation with the present technology.
We carried out a study of quantum pumps in both adiabatic and non-adiabatic
regimes in graphene monolayer reported here, which helps to obtain a clear
physical picture about time-dependent tunneling and will be interesting for
the potential application of carbon-based electronic devices.

\begin{figure*}[t]
\centering
\includegraphics [bb=0 0 820 300, width=1.9\columnwidth]{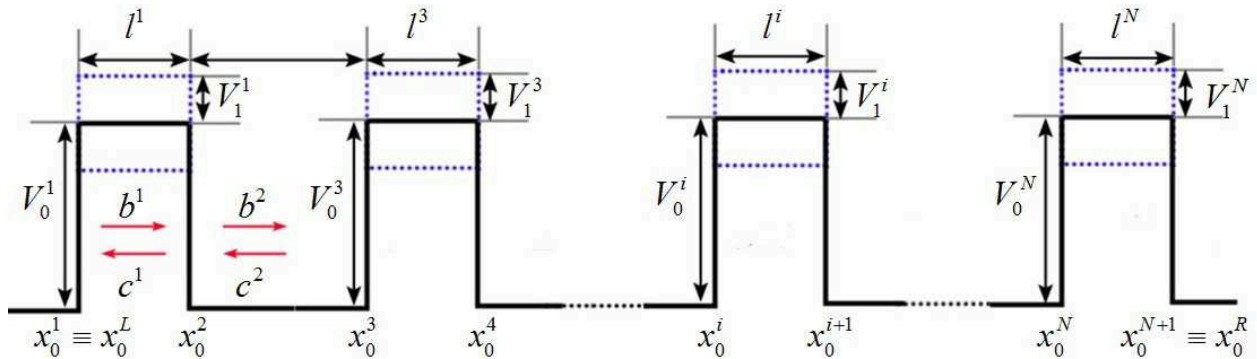}
\caption{Schematic diagram of a series of time-periodic potentials.}
\label{fig:1}
\end{figure*}

In this work, we study theoretically the charge pump by a series of
time-periodic potentials in monolayer graphene with no external bias. We
also find exchange of photons between the oscillating barrier and electrons.
Such processes give rise to transitions from the central band to sidebands,
all of which can convey the electrons and contribute to the pumped current.
The charge pumping is a consequence of the interference of different
propagating modes of which the differences in phase break the time reversal
symmetry.

This paper is organized as follows. In Sec. II, we present the theoretical
model and the calculation method. We show the numerical results and
discussions in Sec. III. Finally, we give the conclusions in Sec. IV.

\section{Theory}

\subsection{Floquet scattering model in graphene \label{sec.IIA}}

We consider electrons transmitting through a series of modulated pumping
potentials in a monolayer graphene sheet. The charge pumping potentials can
be realized by top metallic gates~\cite{Schomerus} or by using the electric
field of a surface acoustic wave.~\cite{Buitelaar} We approximate the
scattering potentials by a sequence of time-periodic rectangular barriers.
The barriers are infinite along the \textit{y}-direction and homogeneous in
each region as illustrated in Fig.~\ref{fig:1}. The potentials are given by:
\begin{equation}
V^{i}(\boldsymbol{r},t)=V_{0}^{i}+V_{1}^{i}\cdot \cos (\omega t+\varphi ^{i})%
\text{, }(x_{0}^{i}<x<x_{0}^{i+1}),
\end{equation}%
where the superscript $i$ indicates the $i$-th region (see Fig.~\ref{fig:1}%
). The height of the barrier is oscillating sinusoidally around $V_{0}^{i}$
with amplitude $V_{1}^{i}$, frequency $\omega $ and phase $\varphi ^{i}$.
Quasiparticles in graphene are formally described by the Dirac-like
Hamiltonian:
\begin{equation}
\hat{H}=\hbar v_{F}\boldsymbol{\sigma }\cdot \boldsymbol{k}+V^{i}(%
\boldsymbol{r},t)\text{.}
\end{equation}%
where $\boldsymbol{k}$ is the momentum, $\sigma _{i}$($i=x,y$) is the
pseudospin Pauli matrix, and the $k$-independent Fermi velocity $v_{F}$
plays the role of the speed of light. \newline
The Floquet theorem asserts that the solution of the eigen-energy problem in
our system has the form:
\begin{equation}
\psi _{F}(\boldsymbol{r},t)=e^{-iE_{F}t/\hbar }\phi (\boldsymbol{r},t),
\end{equation}%
where $E_{F}$ is the Floquet eigen-energy and $\phi (\boldsymbol{r},t)=\phi (%
\boldsymbol{r},t+T)$ is a periodic function, with period~ $T=2\pi /\omega $.
Since the time and position dependence of $V^{i}(\boldsymbol{r},t)$ are separated
in our model, $\phi (\boldsymbol{r},t)$ is separable as $\phi (r,t)=g(%
\boldsymbol{r})f(t)$. This leads to the following equations for $g(%
\boldsymbol{r})$ and $f(t)$, respectively:
\begin{equation}
-i\hbar v_{F}\boldsymbol{\sigma }\cdot \boldsymbol{\nabla }g(\boldsymbol{r}%
)+V_{0}^{i}g(\boldsymbol{r})=Eg(\boldsymbol{r}),
\end{equation}%
\begin{equation}
i\hbar \frac{\partial }{\partial t}f(t)-V_{1}^{i}\cos (\omega t+\varphi
^{i})f(t)=(E-E_{F})f(t),  \label{eq:ft}
\end{equation}%
where $E$ is a constant. Integrating Eq.~(\ref{eq:ft}) gives:
\begin{eqnarray}
f(t) &=&e^{-i(E-E_{F})t/\hbar }\exp (-\frac{i}{\hbar }\int_{0}^{t}V_{1}^{i}%
\cos (\omega t^{^{\prime }}+\varphi ^{i})dt^{^{\prime }})  \notag \\
&=&e^{-i(E-E_{F})t/\hbar }\sum_{n=-\infty }^{n=\infty }J_{n}(\frac{V_{1}^{i}%
}{\hbar \omega })e^{-in\omega t}e^{-in\varphi ^{i}},\ \ \ \ \ \ \
\label{eq:ft2}
\end{eqnarray}%
where we have taken $f(0)=1$, and $J_{n}$ is the Bessel function of the
first kind in order $n$. Note that $f(t)=f(t+T)$, Eq.~(\ref{eq:ft2})
requires that $E=E_{F}+m\hbar \omega $, where $m$ is an integral index of
the sidebands. The translational invariance along the \textit{y} direction
gives rise to the conservation of $k_{y}$, and thus the solutions can be
written as $g(\boldsymbol{r})=g(x)e^{ik_{y}y}$. The equation for $g(%
\boldsymbol{r})$ has a solution:
\begin{eqnarray}
g(\boldsymbol{r}) &=&\sum_{m=-\infty }^{m=\infty }[b_{m}^{i}(%
\begin{array}{c}
1 \\
\frac{k_{m}^{i}+ik_{y}}{E_{m}^{i}}%
\end{array}%
)e^{ik_{m}^{i}(x-x_{0}^{i})}  \notag \\
&&+c_{m}^{i}(%
\begin{array}{c}
1 \\
-\frac{k_{m}^{i}-ik_{y}}{E_{m}^{i}}%
\end{array}%
)e^{-ik_{m}^{i}(x-x_{0}^{i})}]e^{ik_{y}y},  \label{eq:g}
\end{eqnarray}%
where $k_{m}^{i}=sgn(E_{m}^{i})\sqrt{(E_{m}^{i})^{2}-(k_{y})^{2}}$ is the
wavevector of the $n$-th sideband in the $x$-direction, $\varphi ^{i}$ is
the phase of the oscillating potential in the $i$-th region. Note that, here
$E_{m}^{i}\equiv (E_{F}-V_{0}^{i}+m\hbar \omega )/(\hbar v_{F})$ has the
same dimension as the wavevectors $k_{m}^{i}$ and $k_{y}$. According to Eq.~%
\ref{eq:g}, the Floquet energy $E_{F}$ can be determined up to an arbitrary
integer multiplied by $\hbar \omega $. We can define $E_{0}\equiv E_{F}$, $%
E_{n}\equiv E_{F}+n\hbar \omega $. By combining the the solutions for $f(t)$
and $g(\boldsymbol{r})$, we can write the solution for the Floquet state as:
\begin{eqnarray}
\phi _{i}(x,t) &=&\sum_{m=-\infty }^{m=\infty }\sum_{n=-\infty }^{n=\infty
}[b_{m}^{i}(%
\begin{array}{c}
1 \\
\frac{k_{m}^{i}+ik_{y}}{E_{m}^{i}}%
\end{array}%
)e^{ik_{m}^{i}(x-x_{0}^{i})}  \notag \\
&&+c_{m}^{i}(%
\begin{array}{c}
1 \\
-\frac{k_{m}^{i}-ik_{y}}{E_{m}^{i}}%
\end{array}%
)e^{-ik_{m}^{i}(x-x_{0}^{i})}]\ J_{n-m}(\frac{V_{1}^{i}}{\hbar \omega })
\notag \\
&&\cdot e^{-i(n-m)\varphi ^{i}}e^{-i(E_{0}+n\hbar \omega )t/\hbar
}e^{ik_{y}y}.  \label{eq:wave}
\end{eqnarray}%
Note that, in the free region, i.e., $V_{1}^{i}=0$, $J_{n-m}(0)=\delta
_{m,n} $ and Eq.~(\ref{eq:wave}) can also describe the electronic states
correctly. Here we have obtained the wave functions that valid both in the
free and oscillating regions in a unified form. Then we can use Floquet
scattering-matrix formalism (see Appendix.~\ref{Appendix}) to calculate the
pumped current.

\subsection{Pumped current}

Using the distribution function $f_{\alpha }^{out}(E)$ for outgoing
particles and $f_{\alpha }^{in}(E)$ for incoming ones, we can find the
directed current $I_{\alpha }$ in the lead $\alpha $:
\begin{equation}
I_{\alpha }=\frac{eL_{y}}{\pi h}\int_{-\infty }^{\infty
}\int_{-E/hv_{F}}^{E/hv_{F}}\boldsymbol{d}E\boldsymbol{d}k_{y}\{f_{\alpha
}^{out}(E)-f_{\alpha }^{in}(E)\},  \label{eq:i}
\end{equation}%
where, the $f_{\alpha }^{out}(E)$ can be calculated by using the S-matrix $%
S_{F}$ that we have already obtained in Appendix.~\ref{Appendix}:
\begin{equation}
f_{\alpha }^{out}(E)=\sum_{\beta }\sum_{pro-n}|S_{F,\alpha \beta
}(E,E_{n})|^{2}f_{\beta }^{in}(E_{n}).  \label{eq:fout}
\end{equation}%
The subscript $\emph{pro-n}$ indicates that the second summation in Eq.~(\ref%
{eq:fout}) only includes the propagating states with real wave vectors that
contribute to the current. The bound states near the oscillating scatterer
are neglected for electrons in these states are confined around the
boundaries without transmitting far away. Substituting Eq.~(\ref{eq:fout})
into Eq.~(\ref{eq:i}), we have:
\begin{eqnarray}
I_{\alpha } &=&\frac{eL_{y}}{\pi h}\int_{-\infty }^{\infty
}\int_{-E/hv_{F}}^{E/hv_{F}}\boldsymbol{d}E\boldsymbol{d}k_{y}\sum_{\beta
}\sum_{pro-n}  \notag  \label{eq:PI1} \\
&&{|S_{F,\alpha \beta }(E,E_{n})|^{2}(f_{\beta }^{in}(E_{n})-f_{\alpha
}^{in}(E))}.
\end{eqnarray}

The pumped current at the low temperature limit is:
\begin{eqnarray}
I_{\alpha } &=&\frac{eL_{y}}{\pi h}\int_{E_{Fermi}-N\hbar \omega
}^{E_{Fermi}+N\hbar \omega }\int_{-E/hv_{F}}^{E/hv_{F}}\boldsymbol{d}E%
\boldsymbol{d}k_{y}\sum_{\beta }\sum_{pro-n}  \notag \\
&&{|S_{F,\alpha \beta }(E,E_{n})|^{2}(f_{\beta }^{in}(E_{n})-f_{\alpha
}^{in}(E))},  \label{eq:PI}
\end{eqnarray}%
where, $\alpha $ is the outgoing terminal, $\beta $ is the incoming
terminal, $N$ is the maximum index of the sidebands that need to be included
in the calculation and it is determined by the amplitude and frequency of
the oscillation (proportional to $V_{1}/\hbar \omega $). \ Next we consider
the adiabatic limitation: $\omega \rightarrow 0$. In this situation, the
Floquet scattering matrix is almost energy independent when $E$ changes by $%
N\hbar \omega $. We can rewrite the Floquet scattering matrix in adiabatic
approximation as ${S_{F,\alpha \beta }(E,E_{n})}\rightarrow $ ${S_{0,\alpha
\beta ,n}(E)}$, and the difference of the Fermi distribution as ${%
f(E_{n})-f(E)=}n\hbar \omega \frac{{f(E_{n})-f(E)}}{{E_{n}-E}}\rightarrow
n\hbar \omega \frac{\partial {f_{0}(E)}}{\partial E}$, where $n$ denotes the $n$-th sideband in which the electrons are scattered to. Inserting the above
expressions into Eq.~(\ref{eq:PI1}), the pumped current in the adiabatic
pumping regime can be reduced to,
\begin{eqnarray}
I_{\alpha } &=&\frac{eL_{y}\omega }{2\pi ^{2}}\int_{-\infty }^{\infty
}\int_{-E/hv_{F}}^{E/hv_{F}}\boldsymbol{d}E\boldsymbol{d}k_{y}\sum_{\beta
}\sum_{pro-n}  \notag \\
&&n{|S_{0,\alpha \beta ,n}(E)|^{2}}\frac{\partial {f_{0}(E)}}{\partial E}.
\label{eq:PIad}
\end{eqnarray}%
We note en passant that, at the low temperature limit, only the particles
close to the Fermi level can contribute to the adiabatic pumped current as
indicated by the derivative of the Fermi distribution in Eq.~(\ref{eq:PIad}%
). We can also use the scattering matrix to calculate some other important
physical quantities such as the directed heat current, shot noise, Wigner
delay time.~\cite{Zhu,Li,Moskalets,Moskalets2}

\section{Numerical Results and Discussions}

\subsection{Transmission asymmetry}

In this section, we use the above formalism to investigate quantum tunneling
through double oscillating barriers schematically illustrated in Fig.~\ref%
{fig:1} with only the first two oscillating barriers included. It is clear
that a single time-periodic barrier with time-reversal symmetry can not
produce a pumped current. Here we consider only two oscillating potentials
which are sufficient to present a clear physical picture. The heights of the
two barriers oscillate in time with the same frequency $\omega $, amplitude $%
V_{1}$, but with a phase lag $\Delta \varphi $. Thus for a pump with $\Delta
\varphi \neq n\pi $ (n is a integer), the time-reversal symmetry is broken
and a nonzero pumped current can be produced. To understand the origin of
the directed current when the chemical potentials at both sides of the
scatterers are equal, we first examine the transmission asymmetry for incoming
waves in a certain mode $E_{0}=E$ from both sides of the scattering region.

\begin{eqnarray}
T_{net} &=&\sum_{pro-n}\{|S_{F,\alpha \beta }(E_{n},E)|^{2}\frac{%
v_{n}^{\alpha }}{v_{0}^{\beta }}f_{\beta }^{in}(E)  \notag \\
&&-|S_{F,\beta \alpha }(E_{n},E)|^{2}\frac{v_{n}^{\beta }}{v_{0}^{\alpha }}%
f_{\alpha }^{in}(E)\},  \label{eq:netT}
\end{eqnarray}%
where $\alpha $ denotes the right terminal and $\beta $ denotes the left
terminal.

\begin{figure}[h]
\centering
\includegraphics [width=\columnwidth]{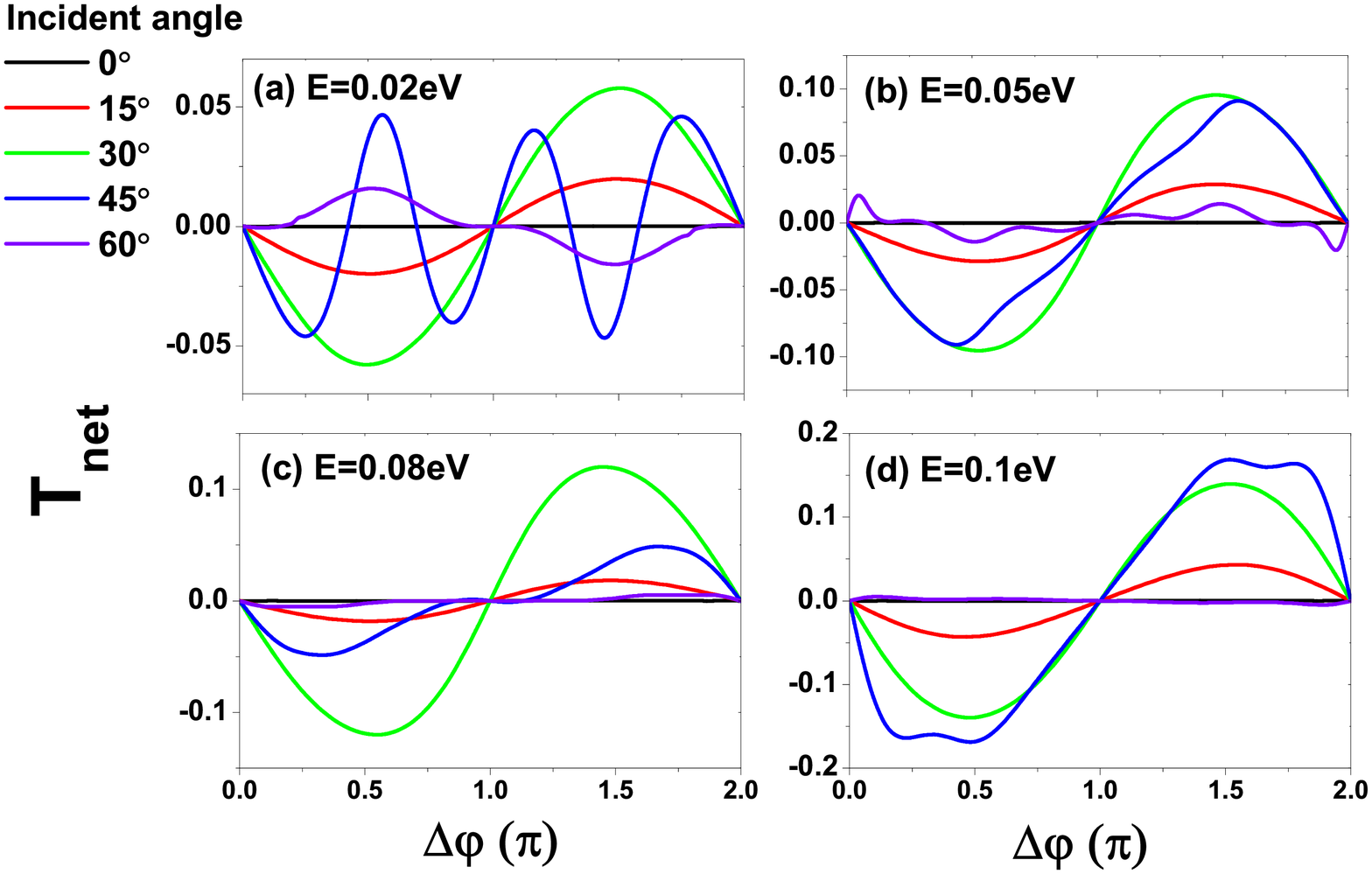}
\caption{The transmission asymmetry as a function of the phase lag $%
\Delta\protect\varphi$ at different incident angles and energies, with $%
V_{0}=200$ meV, $V_{1}=9$ meV, $\hbar \protect\omega=3$ meV, $l_{1}=l_{3}=50$
nm $l_{2}=25$ nm}
\label{fig:2}
\end{figure}

\begin{figure}[t]
\centering
\includegraphics [width=\columnwidth]{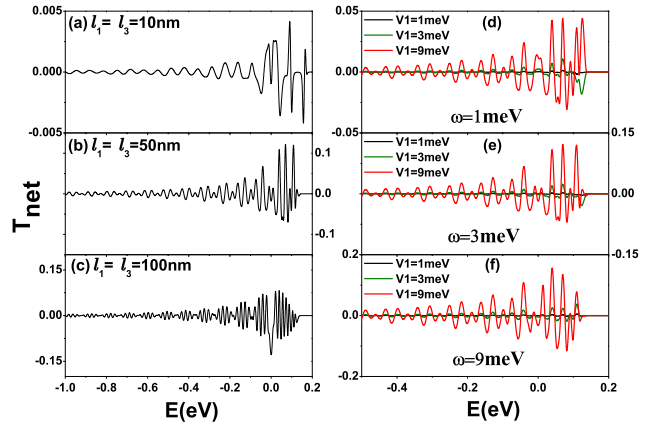}
\caption{(a)-(c) The transmission asymmetry $T_{net}$ as a function of incident
energy with different barrier widths. $V_{0}=200$ meV, $V_{1}=9$ meV, $\hbar
\protect\omega =3$ meV. (d)-(f) The transmission asymmetry $T_{net}$ as a function
of incident energy with different amplitudes and frequency of the
oscillating potentials. $l_{1}=l_{3}=50$ nm, $V_{0}=200$ meV. In all panels
the incident angle and the distance between the two barriers are fixed,
i.e., $\protect\phi=30^{\circ}$, $l_{2}=25$ nm.}
\label{fig:3}
\end{figure}

In Fig.~\ref{fig:2}, we plot the transmission asymmetry probability as a function
of the phase lag $\Delta \varphi $ at different incident angles and
energies. Anisotropic tunneling behavior is observed in this time-periodic
case. Note that for normal incidence, $T_{net}$ is always zero. This
phenomena is the result of Klein tunneling i.e., the incident electrons at
both sides of the scatterer can perfectly transmit to the opposite side
resulting in the vanish of the transmission asymmetry. In Fig.~\ref{fig:3}(a)-(c),
we present the transmission asymmetry probability as a function of incident energy
under a certain Fermi energy $E_{F}=0.2$ eV, while the widths of the
barriers are different. Our calculation shows that $T_{net}$ vanishes
gradually as the incident energy decreases, since the influence of the
oscillating potentials that break the time-reversal symmetry become
negligible when the incident energy is far bellow $V_{0}$ as well as the
oscillating potentials. The transmission asymmetry probability oscillates more
seriously when the barrier is thicker by comparing Fig.~\ref{fig:3}(a)-(c).
This is related to the interference between the multireflections of the
electron waves inside the barriers. So it shows more pronounced resonant
tunneling behaviors as the barrier width increases. We also investigate the
effect of the oscillating amplitude $V_{1}$ and frequency $\omega $ on $%
T_{net}$ as shown in Fig.~\ref{fig:3}(d)-(f). We find that the transmission asymmetry probability can be enlarged both by increasing $V_{1}$ or $%
\omega $ for small values of $\omega $. Such results are obvious since
larger $V_{1}$ gives rise to a greater number of sidebands serving as the
transmission modes, while larger $\omega $ gives rise to larger modulation
quantum of energy that an electron exchanged with the oscillating potentials
during the tunneling processes. In Fig.~\ref{fig:3}, we find that the transmission asymmetry vanishes when the incident energy approaches $V_{0}$. This is
because the \textit{x}-component of the wave vector $k_{m}$ of the central
band and first few sidebands in the barrier are imaginary in the incident
energy range closed to $V_{0}$. Imaginary wave vectors corresponds to
evanescent modes, that decay exponentially and thus lead to total reflection
at both sides of the scatterer.~\cite{Wu} The breaking of the time-reversal
symmetry by the interference of different traversal modes (i.e., central
band and sidebands) is attributed to the phase lag between the two
oscillating potentials.

\subsection{Adiabatic Pumped Current}

\begin{figure}[tbp]
\centering
\begin{minipage}[c]{0.9\columnwidth}
\centering
\includegraphics[width=\columnwidth]{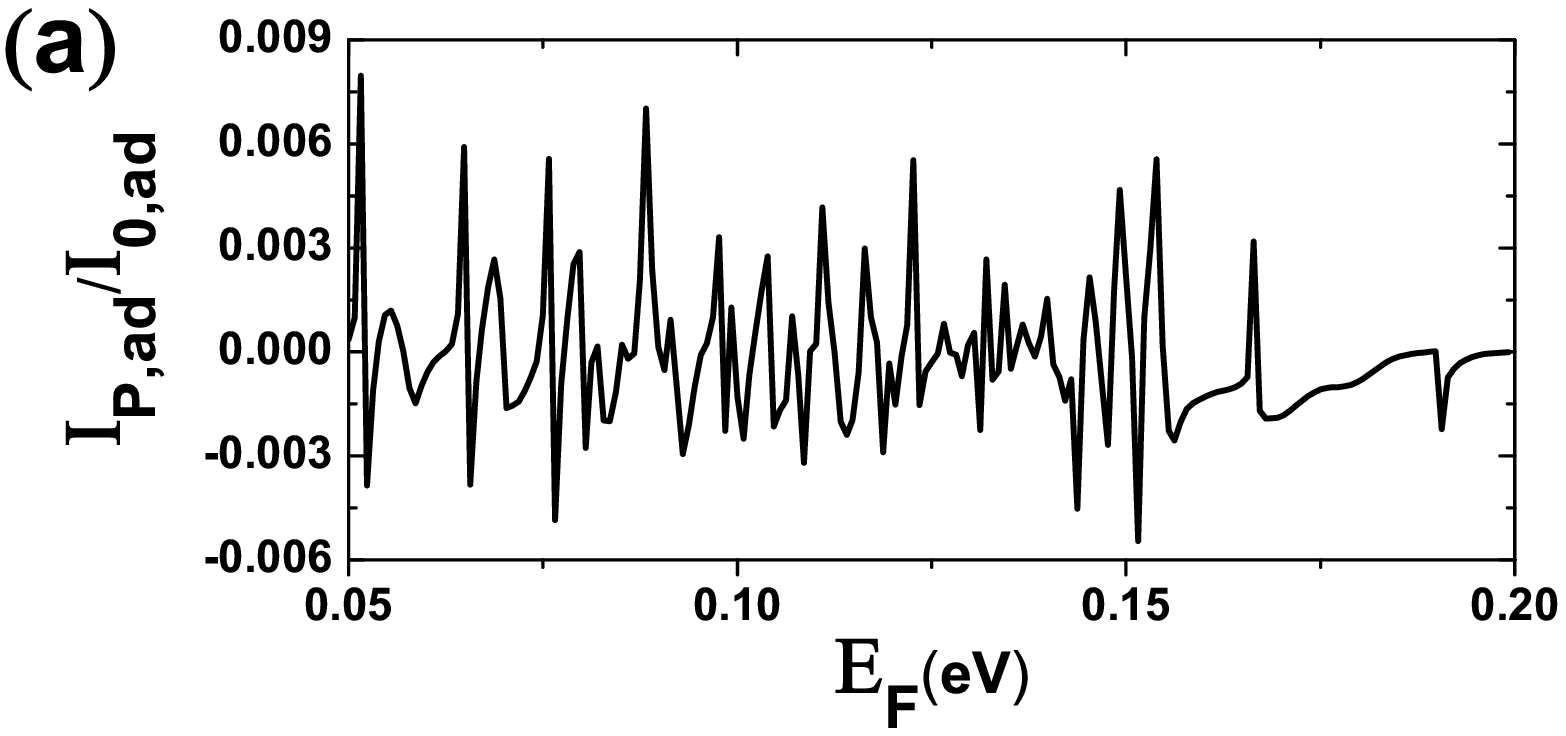}
\end{minipage}\newline
\begin{minipage}[c]{0.9\columnwidth}
\centering
\includegraphics[width=\columnwidth]{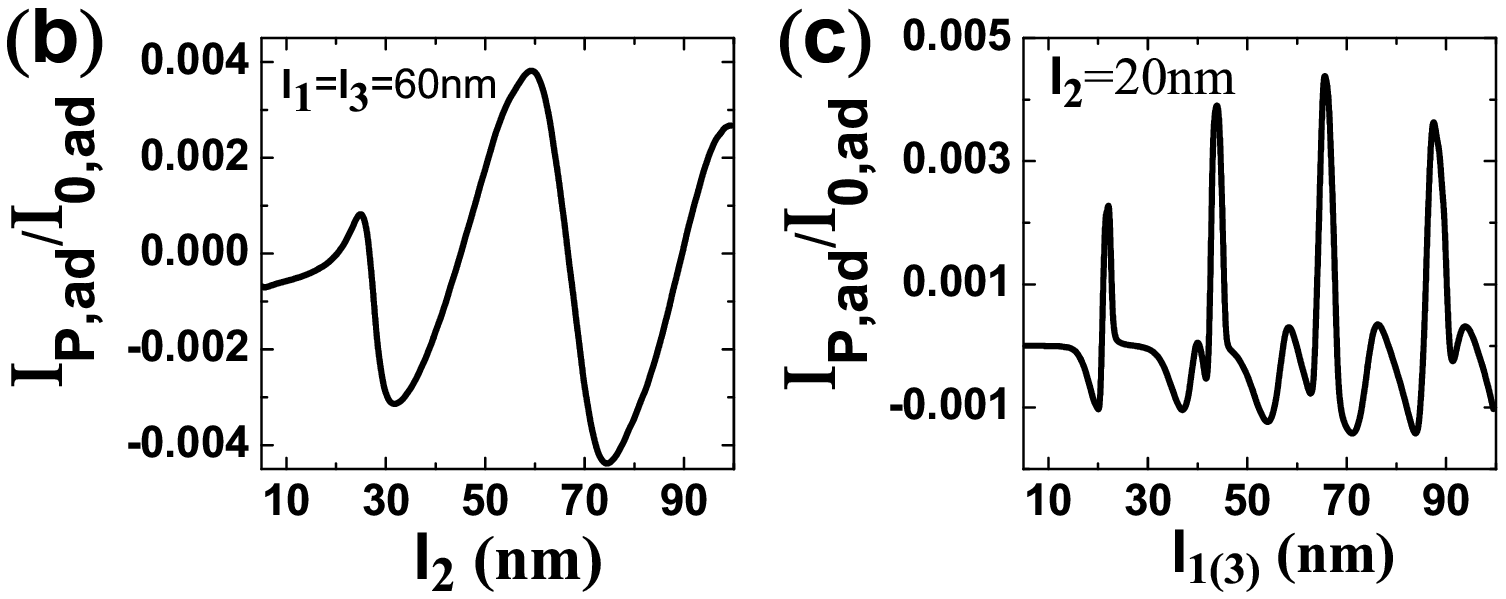}
\end{minipage}
\caption{(a) Energy dependence of the adiabatic pumped current, with $\Delta%
\protect\varphi=\protect\pi/2$, $\protect\omega=0.03$ meV, $V_{1}=3\protect%
\omega$, $V_{0}=0.2$ eV $l_{2}=60$ nm, $l_{1}=l_{3}=100$ nm, and $%
I_{0,ad}\equiv eL_{y}\protect\omega/(2\protect\pi ^{2})$. (b)-(c) The
adiabatic pumped currents versus the barrier separation and barrier width
for fixed Fermi energy $E_{F}=80$ meV.}
\label{fig:4}
\end{figure}

Nonzero transmission asymmetry can give rise to a finite charge current. Next we
proceed to calculate the pumped current in adiabatic limit: $\omega
\rightarrow 0$. The expression of adiabatic pumped current in Eq.~(\ref%
{eq:PIad}) is valid when the Floquet scattering matrix is energy independent
on the scale of the order of $N\hbar \omega$. The lager the number $N$ of
the sidebands included, the smaller the frequencies for which the adiabatic
approximation is valid. As discussed above, $N$ is determined by the ratio
of the oscillation amplitude to the frequency. In this subsection, we
consider a weak pump with $N=3$. For a strong pump, $N$ is much larger and
the adiabatic approximation is valid only at smaller frequencies. In Fig.~%
\ref{fig:3}, we can see that the transmission asymmetry shows a resonance-like
behavior of the order of the width $\delta E \sim 0.02$ eV resonance,
indicating that the Floquet scattering matrix changes significantly over
this energy scale. Therefore we can choose an oscillating frequency of $%
\omega =0.03$ meV, for which the adiabatic approximation requirement $N\hbar
\omega \ll \delta E$ is satisfied. Note that at zero temperature the pumped
current is dominated by the electrons at the Fermi energy. The pumped
current $I_{P,ad}$ shows similar resonance-like behaviors as transmission asymmetry as expected (see Fig.~\ref{fig:4}(a)). $I_{P,ad}$ oscillates as
a function of the Fermi energy $E_{F}$, becomes smooth and eventually
vanishes when the fermi energy approaches the static potential height $V_{0}$%
. It is much different from that of a conventional quantum pump based on
usual nano-scale semiconductors. This remarkable feature can be attributed
to the same reason as the vanished transmission asymmetry when $E_{F}$ is close to
$V_{0}$ as we examined in the last subsection, i.e., the transmission
probability in graphene through a high barrier is lager than that of a low
barrier which is closed to the incident energy. Another difference between
the quantum pumps in graphene and that in the conventional two-dimensional
electron gas system is that the pumped current in graphene can change from
positive to negative or reversely as the energy increases for a constant
phase lag, while the direction of conventional pumped current is determined
by the phase lag of the to oscillating potentials.~\cite{RZhu} We also
investigated the influence of the geometrical sizes of the oscillating
double barriers. In Figs.~\ref{fig:4}(b) and (c), we plot the dependence of
adiabatic pumped currents both on the barrier separation $l_{2}$ and width $%
l_{1(3)}$. The adiabatic pumped current also oscillates when the barrier
separation or width are changed. This is caused by the Fabry-P\'{e}rot
resonant modes formed between the interfaces due to the multiple
reflections.~\cite{Torres2} These results agree with the previous work on
adiabatic pumping in graphene, e.g. see~\cite{RZhu}. With increasing pumped
frequency, the deviation of the actual pumped current calculated by Eq.(\ref%
{eq:PI}) from the adiabatic one calculated by Eq.(\ref{eq:PIad}) becomes
significant, and thus the adiabatic approximation of Eq.(\ref{eq:PIad}) as
well as the parametric scattering approach used in Ref.\cite{RZhu} are not
valid anymore.

\subsection{Non-adiabatic Pumped Current}

\begin{figure}[tbp]
\centering
\begin{minipage}[c]{0.7\columnwidth}
\centering
\includegraphics[width=\columnwidth]{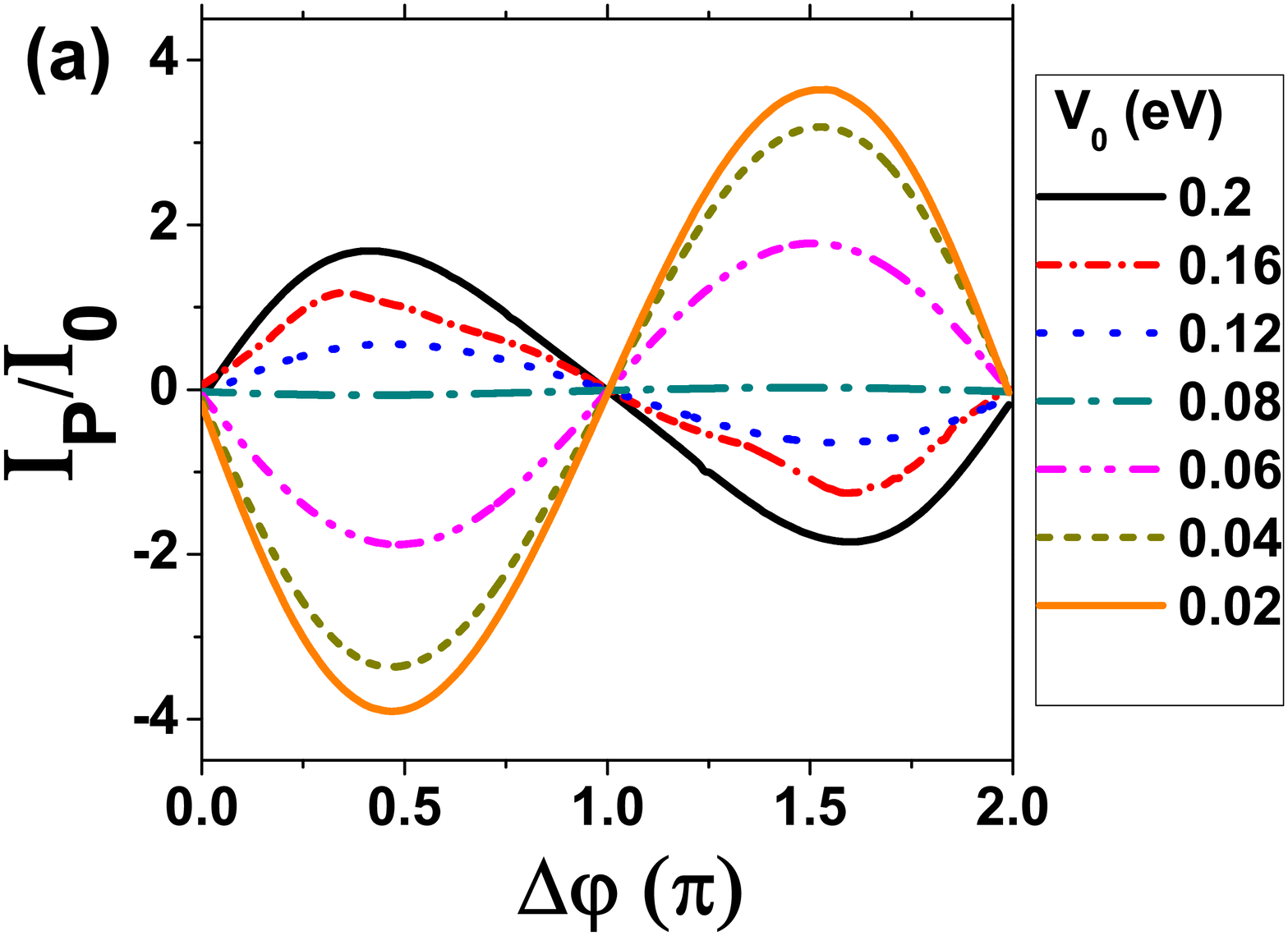}
\end{minipage}%
\begin{minipage}[c]{0.3\columnwidth}
\centering
\includegraphics[bb=0 0 820 1430, width=\columnwidth]{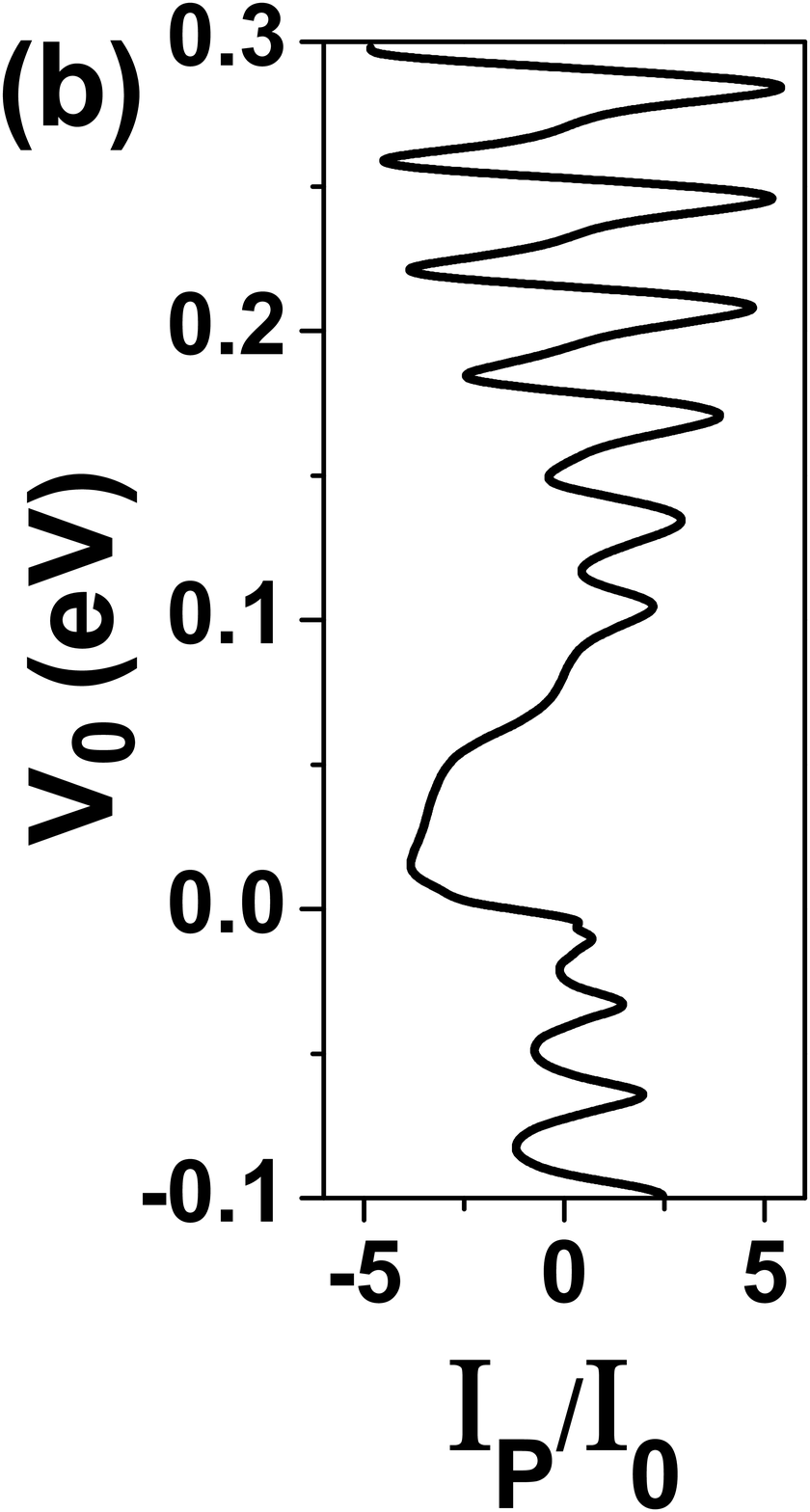}
\end{minipage}
\caption{(a) The pumped currents versus the phase lag $\Delta\protect\varphi
$ for different $V_{0}$, with $E_{F}=80$ mev, $\protect\omega=3$ meV, $%
V_{1}=9$ meV, $I_{0}\equiv eL_{y}/(10^{5}\protect\pi h)$. (b) The pumped
currents versus $V_{0}$, with $\Delta\protect\varphi=\protect\pi/2$.}
\label{fig:5}
\end{figure}

Next, we analyze the behavior of non-adiabatic pumped current $I_{P}$ as a
function of phase lag $\Delta \varphi $ by using Eq.~(\ref{eq:PI}). The
incident electrons are homogeneous in direction under the same Fermi energy $%
E_{F}$ from both sides of the scatterer. Fig.~\ref{fig:5}(a) shows the
general characteristics of quantum pumping in graphene. The oscillating
frequency is set to $\omega=3$ meV, and thus $N\hbar\omega$ is of the same
order of $\delta E$. The adiabatic approximation discussed in last
subsection is no longer valid. The sizes of the double oscillating barriers
are fixed for $l_{2}=25$ nm, $l_{1}=l_{3}=50$ nm in the calculation of this
subsection. The numerical results show that the pumped current $I_{P}$ is
sinusoidal or cosinoidal dependence on phase lag $\Delta \varphi $. We find
that the static barrier $V_{0}^{i}$ can also effectively control the
amplitude and direction of the current as shown in Fig.~\ref{fig:5}(b). The
height of the static barrier $V_{0}^{i}$ plays an important role in
determining the pumping, since it is related to the component of the wave
vector along the x-direction in the barrier region. For electron in the $n$%
-th energy sideband $|E+n\hbar \omega -V_{0}^{i}|<\hbar v_{F}k_{y}$, the
corresponding \textit{x}-component of the wave vector $k_{n}^{i}$ becomes
imaginary. Such electrons are in evanescent sidebands and do not directly
contribute to the current. If the static barrier becomes higher, the
electronic states outside the barrier match the hole states inside the
barrier. Then the transmission modes in the barrier come back to propagating
states, resulting in an increasing overall pumped current amplitude. This
phenomena in graphene is different from the transmission in ordinary
semiconductor two dimensional electron gas, while there is no available hole
state inside the barrier contributing to ordinary tunneling. One can see in
Fig.~\ref{fig:5}(b) that the pumped current oscillates when the static
barrier potential is increased. This is caused by the Fabry-P\'{e}rot
resonant modes formed in the double barrier structure due to the multiple
reflections.~\cite{Torres2}

\begin{figure}[tbp]
\centering
\begin{minipage}[c]{0.5\columnwidth}
\centering
\includegraphics[width=\columnwidth]{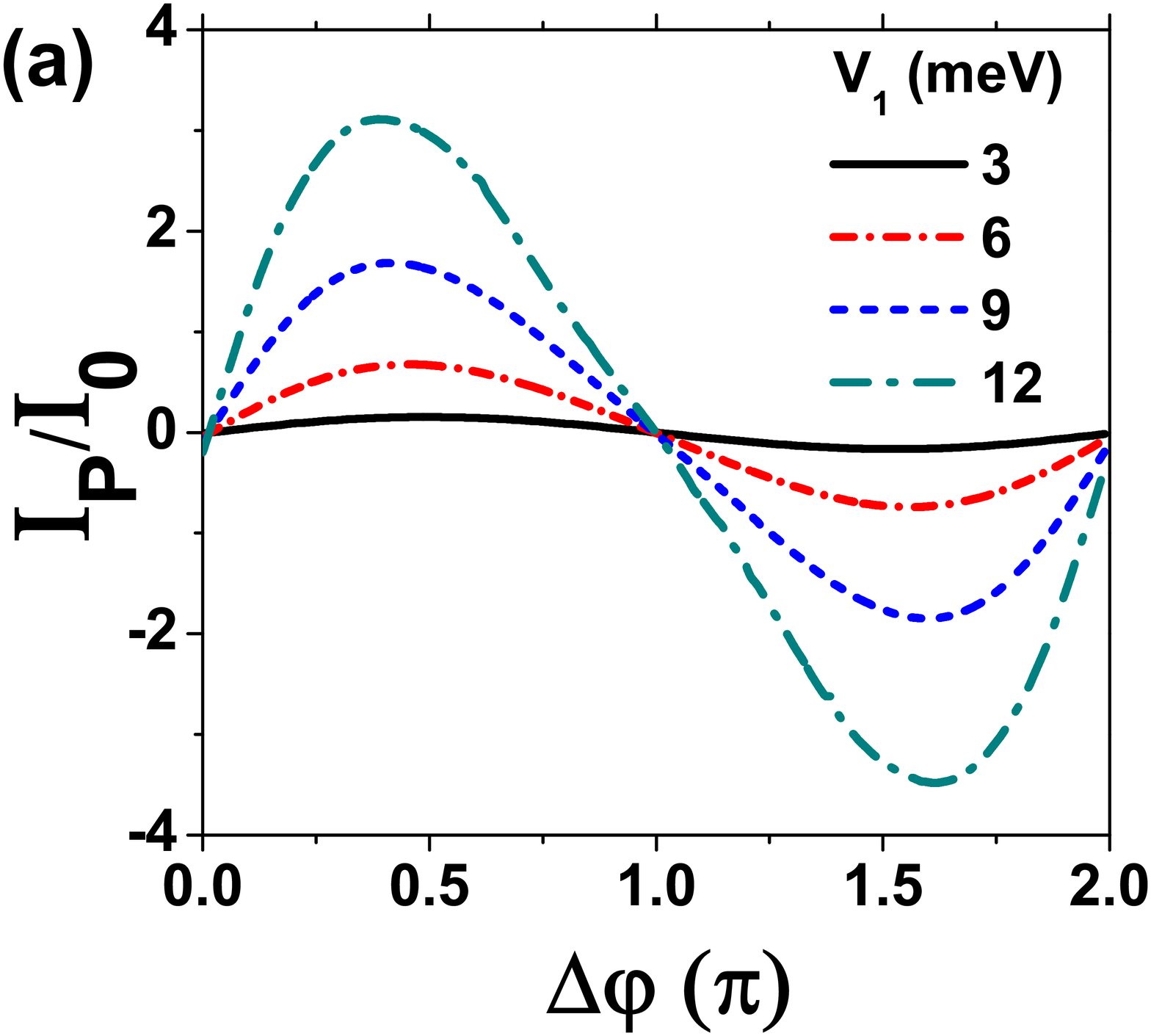}
\end{minipage}%
\begin{minipage}[c]{0.5\columnwidth}
\centering
\includegraphics[ width=\columnwidth]{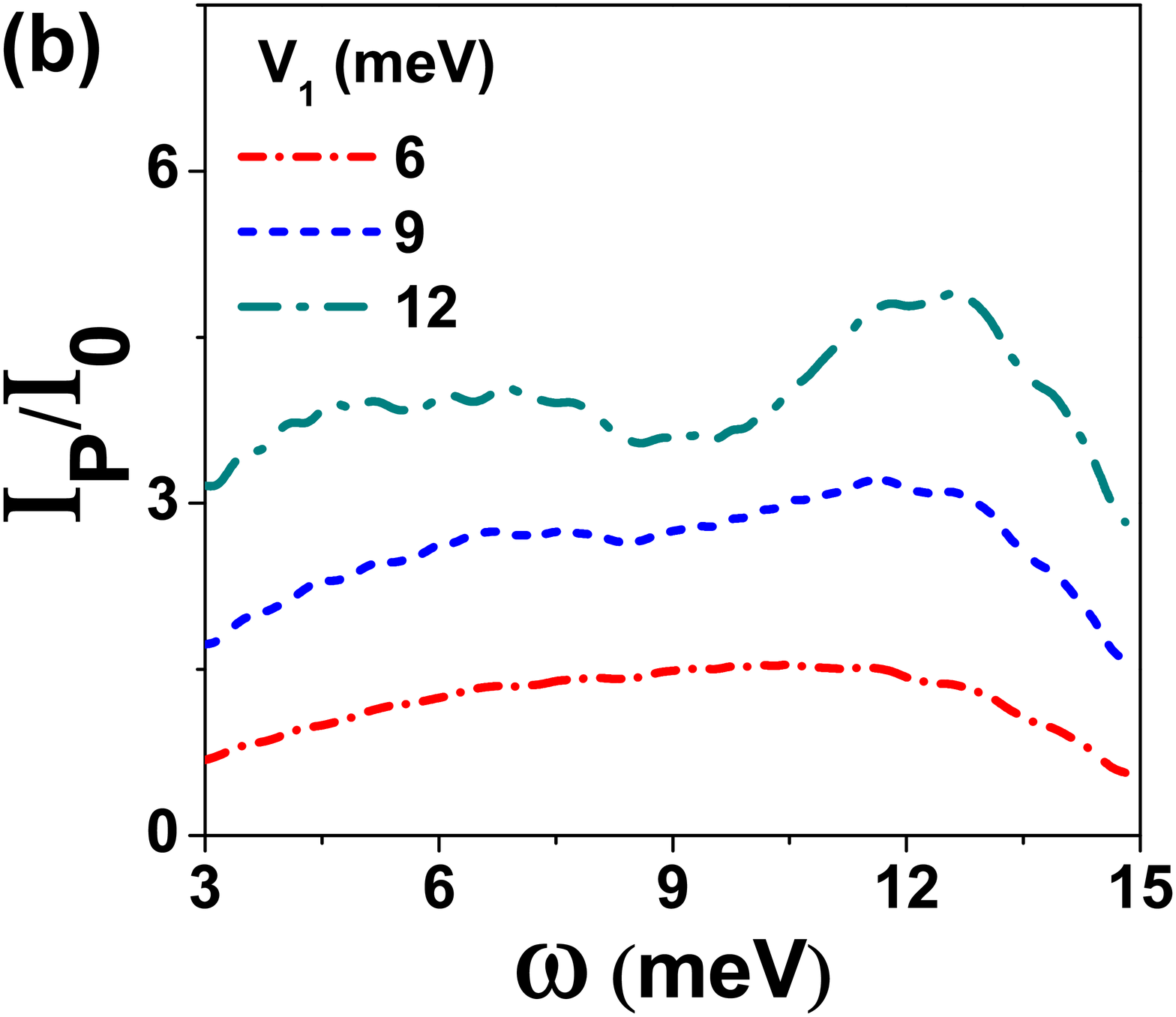}
\end{minipage}
\caption{(a) The pumped currents versus the phase lag $\Delta\protect\varphi
$ for different $V_{1}$, with $E_{F}=80$ mev, $\protect\omega=3$ meV, $%
V_{0}=200$ meV, $I_{0}\equiv eL_{y}/(10^{5}\protect\pi h)$. (b) The pumped
currents versus $\protect\omega$, with $\Delta\protect\varphi=\protect\pi/2$%
. }
\label{fig:6}
\end{figure}

The pumped current can also be tuned by the oscillating amplitude $V_{1}$
and frequency $\omega $ as shown in Fig.~\ref{fig:6}. We find that the
pumped current increases with the oscillating amplitude $V_{1}$. This is
because a larger $V_{1}$ can transfer electrons to higher sidebands and thus
increase the transmission modes of which the interferences become more
intensive. This behavior is much different from that of an adiabatic pump.
In the limit of $\omega \rightarrow 0$, the adiabatic pumped current is
irrespective of the amplitude of the oscillating potentials $V_{1}$, however the criterion of adiabatic approximation depends
strongly on $V_{1}$.~\cite{Moskalets2}

In Fig.~\ref{fig:6}(b), we inspect the influence of the frequency $\omega $
for several fixed amplitude $V_{1}$. The pumped current shows a trend of
rise first, then fall after $\omega $ exceeding $V_{1}$. For small $\omega $%
, electrons can gain a larger modulation energy quantum from the oscillating
potentials with increasing $\omega $. The interference of sidebands with
higher energies gives rise to a lager pumped current. With increasing pumped
frequency $\omega$ under a constant oscillating potential $V_{1}$, the
oscillating double barriers will transform from a strong pump to a weak
pump. In the case that $\omega $ is lager than $V_{1}$, the oscillating pump
is too weak to lift the electrons from the central band to any sidebands and
thus leads to the drop of the pumped current. This feature dose not exist in
adiabatic pumping regimes, since the pumped frequency $\omega$ is always
much smaller than the oscillating potential $V_{1}$, and the adiabatic
current always increases monotonely with $\omega$.~\cite{Agarwal} Therefore,
the non-adiabatic pumped current can be tuned by changing the phase lag,
barrier height, oscillating amplitude and frequency.

\section{Conclusion}

In summary, we applied the iterative Floquet scattering-matrix method to the
Dirac system, which is able to calculate the pumped current induced by an
arbitrary number of parametric oscillating potentials between two electron
reservoirs without external bias. Based on this approach, we investigate
theoretically the quantum pumping driven by a series of time-periodic
potentials in graphene monolayer both in adiabatic and non-adiabatic
regimes. We have determined how the pumped current depends on the phase lag, height
of the barrier, frequency and amplitude of the oscillation. Finally, the
physical mechanism of quantum pumping is attributed to the interference of
different central band and sidebands of which the phases factor break the
time reversal symmetry.

\begin{acknowledgments}
The work described in this paper was supported by the Collaborative Research
Fund of the Research Grants Council of Hong Kong SAR, China under Project
No. CityU5/CRF/08 and NSFC Grant No.11104232.
\end{acknowledgments}

\appendix

\section{Transfer-matrix and Scattering-matrix \label{Appendix}}

In Sec.~\ref{sec.IIA}, we have obtained the wave functions for all the free
and oscillating regions. The wave functions (see Eq.~(\ref{eq:wave})) must
be continuous at each boundary $x=x_{0}^{0,1,2...}$. At the boundary between
region $i$ and $i+1$ ($x=x_{0}^{i+1}$), the coefficients matrices can be
related by the transfer matrix $\boldsymbol{M}(i,i+1)$,
\begin{equation}
\dbinom{B^{i}}{C^{i}}=\boldsymbol{M}(i,i+1)\dbinom{B^{i+1}}{C^{i+1}},
\label{eq:TM}
\end{equation}%
where $\boldsymbol{B}$ ($\boldsymbol{C}$) is the probability amplitudes
column vector of the right(left)-going waves (see Fig.~\ref{fig:1}).
Matching the wavefunctions at $x_{0}^{i+1}$ and realizing that ${%
e^{-i(E_{0}+n\hbar \omega )t/\hbar }}$ with different $n$ are orthogonal, we
have:
\begin{eqnarray}
&&\sum_{n}\sum_{m}J_{n-m}(\frac{V_{1}^{i}}{\hbar \omega })\cdot
e^{-i(n-m)\varphi
^{i}}[b_{m}^{i}e^{ik_{m}^{i}l^{i}}+c_{m}^{i}e^{-ik_{m}^{i}l^{i}}]  \notag \\
&=&\sum_{n}\sum_{m}J_{n-m}(\frac{V_{1}^{i+1}}{\hbar \omega })\cdot
e^{-i(n-m)\varphi ^{i+1}}[b_{m}^{i+1}+c_{m}^{i+1}],  \label{eq:boundary1}
\end{eqnarray}%
and
\begin{eqnarray}
&&\sum_{n}\sum_{m}J_{n-m}(\frac{V_{1}^{i}}{\hbar \omega })\cdot
e^{-i(n-m)\varphi ^{i}}\ \ \ \ \ \ \ \ \ \ \ \ \ \ \ \ \ \ \ \ \ \   \notag
\\
&&\cdot \lbrack \frac{k_{m}^{i}+\boldsymbol{i}k_{y}}{E_{m}^{i}}\cdot
b_{m}^{i}e^{ik_{m}^{i}l^{i}}-\frac{k_{m}^{i}-\boldsymbol{i}k_{y}}{E_{m}^{i}}%
\cdot c_{m}^{i}e^{-ik_{m}^{i}l^{i}}]  \notag \\
&=&\sum_{n}\sum_{m}J_{n-m}(\frac{V_{1}^{i+1}}{\hbar \omega })\cdot
e^{-i(n-m)\varphi ^{i+1}}\ \ \ \ \ \ \ \ \ \ \ \ \ \ \   \notag \\
&&\cdot \lbrack \frac{k_{m}^{i+1}+\boldsymbol{i}k_{y}}{E_{m}^{i+1}}\cdot
b_{m}^{i+1}-\frac{k_{m}^{i+1}-\boldsymbol{i}k_{y}}{E_{m}^{i+1}}\cdot
c_{m}^{i+1}],  \label{eq:boundary2}
\end{eqnarray}%
where, $l^{i}=x_{0}^{i+1}-x_{0}^{i}$, is the distance of the $i$-th region.
In order to convert the above two equations into the matrix form, we define:
\begin{equation}
(\gamma ^{i})_{nm}\equiv e^{ik_{m}^{i}l^{i}}\cdot \delta _{mn}\text{;}
\end{equation}%
\begin{equation}
(P^{i})_{nm}\equiv J_{n-m}(\frac{V_{1}^{i}}{\hbar \omega })\cdot
e^{-i(n-m)\varphi ^{i}}\text{;}
\end{equation}%
\begin{equation}
(Q_{R}^{i})_{nm}\equiv J_{n-m}(\frac{V_{1}^{i}}{\hbar \omega })\cdot
e^{-i(n-m)\varphi ^{i}}\cdot (\frac{k_{m}^{i}+\boldsymbol{i}k_{y}}{E_{m}^{i}}%
)\text{;}
\end{equation}%
\begin{equation}
(Q_{L}^{i})_{nm}\equiv J_{n-m}(\frac{V_{1}^{i}}{\hbar \omega })\cdot
e^{-i(n-m)\varphi ^{i}}\cdot (-\frac{k_{m}^{i}-\boldsymbol{i}k_{y}}{E_{m}^{i}%
})\text{.}
\end{equation}%
Using the boundary conditions Eq.~(\ref{eq:boundary1}) and Eq.~(\ref%
{eq:boundary2}) we can obtain the transfer matrix:
\begin{eqnarray}
\boldsymbol{M}(i,i+1) &=&\left(
\begin{array}{cc}
\gamma ^{i} & 0 \\
0 & (\gamma ^{i})^{-1}%
\end{array}%
\right) ^{-1}\left(
\begin{array}{cc}
P^{i} & P^{i} \\
Q_{R}^{i} & Q_{L}^{i}%
\end{array}%
\right) ^{-1}  \notag \\
&&\cdot \left(
\begin{array}{cc}
P^{i+1} & P^{i+1} \\
Q_{R}^{i+1} & Q_{L}^{i+1}%
\end{array}%
\right) .
\end{eqnarray}

If we define:
\begin{equation}
T(i,i+1)\equiv \left(
\begin{array}{cc}
P^{i} & P^{i} \\
Q_{R}^{i} & Q_{L}^{i}%
\end{array}%
\right) ^{-1}\left(
\begin{array}{cc}
P^{i+1} & P^{i+1} \\
Q_{R}^{i+1} & Q_{L}^{i+1}%
\end{array}%
\right) ,
\end{equation}%
we have all the four sub-transfer matrix:
\begin{eqnarray}
M_{11}(i,i+1) &=&(\gamma ^{i})^{-1}T_{11}(i,i+1),  \notag \\
M_{12}(i,i+1) &=&(\gamma ^{i})^{-1}T_{12}(i,i+1),  \notag \\
M_{21}(i,i+1) &=&(\gamma ^{i})T_{21}(i,i+1),  \notag \\
M_{22}(i,i+1) &=&(\gamma ^{i})T_{22}(i,i+1).  \label{eq:tm}
\end{eqnarray}%
So far, we have obtained the transfer matrices of each region. But for an
evanescent mode, where $k_{m}^{i}$ becomes imaginary, the $(\gamma
^{i})_{mm} $ and$(\gamma ^{i})_{mm}^{-1}$ will , respectively, decay or grow
exponentially. An approatch to avoid this problem is using the scattering
matrix formalism. In the following, we obtain the scattering matrix for the
graphene system following Xu's work on quantum antidot arrays.~\cite{Xu}

The scattering matrix $S(L,R)$, connecting the incoming and outgoing
channels, is defined as follows:
\begin{eqnarray}
\dbinom{B^{R}}{C^{L}} &=&S(L,R)\dbinom{B^{L}}{C^{R}}  \notag \\
&\equiv &\left(
\begin{array}{cc}
S_{11}(L,R) & S_{12}(L,R) \\
S_{21}(L,R) & S_{22}(L,R)%
\end{array}%
\right) \dbinom{B^{L}}{C^{R}}.
\end{eqnarray}

Then, we use the iterative scheme, if we already have $S(L,i)$:
\begin{equation}
\dbinom{B^{i}}{C^{L}}=S(L,i)\dbinom{B^{L}}{C^{i}}\text{,}
\end{equation}%
with the help of the transfer matrix (see Eq.~(\ref{eq:TM})), we can obtain
the scattering matrix of the next boundary $S(L,i+1)$:

\begin{eqnarray}
S_{11}(L,i+1) &=&[1-M_{11}^{-1}(i,i+1)S_{12}(L,i)M_{21}(i,i+1)]^{-1}  \notag
\\
&&\cdot M_{11}^{-1}(i,i+1)S_{11}(L,i)  \notag \\
&&  \notag \\
S_{12}(L,i+1) &=&[1-M_{11}^{-1}(i,i+1)S_{12}(L,i)M_{21}(i,i+1)]^{-1}  \notag
\\
&&\cdot M_{11}^{-1}(i,i+1)S_{12}(L,i)M_{22}(i,i+1)  \notag \\
&&-M_{11}^{-1}(i,i+1)M_{12}(i,i+1)]\   \notag \\
&&  \notag \\
S_{12}(L,i+1) &=&S_{22}(L,i)M_{21}(i,i+1)S_{11}(L,i+1)  \notag \\
&&+S_{21}(L,i)  \notag \\
&&  \notag \\
S_{22}(L,i+1) &=&S_{22}(L,i)M_{21}(i,i+1)S_{12}(L,i+1)  \notag \\
&&+S_{22}(L,i)\cdot M_{22}(i,i+1).  \label{eq:sm}
\end{eqnarray}%
We have obtained the iterative relations of the scatter-matrices in Eq. (\ref%
{eq:sm}). It is easy to write the initial S-matrix: $S_{L,L}=\hat{I}$. Then
we can get every desired S-matrix $S_{L,R}$ using the iterative scheme step
by step. Note that the current associated with a scattered wave is
proportional to the square of the wavefunction multiplied by the velocity.
In order to ensure current conservation The unitariy Floquet scattering
matrix by incorporating the velocities of different modes in the incident
and outgoing terminals,
\begin{equation}
S_{F,\alpha \beta }(m,n)=S_{\alpha \beta }(m,n)|^{2}\frac{v_{m}^{\alpha }}{%
v_{n}^{\beta }},
\end{equation}%
where $\alpha $ and $\beta $ denote the terminals, $m$ and $n$ label the
transmission modes. In this research of a series of
time-periodic potentials, the velocities in the propagating direction is
calculated as follows:
\begin{equation}
v_{m}^{\alpha }=\frac{1}{\hbar }\frac{\partial }{\partial k_{m}^{\alpha }}%
(\hbar v_{F}E_{m}^{\alpha }(k_{m}^{\alpha },k_{y}))\ =\frac{%
v_{F}k_{m}^{\alpha }}{E_{m}^{\alpha }}\text{.\ }
\end{equation}%
This iterative scattering-matrix method has the advantage that we can
effectively investigate a complicate system consisting of an arbitrary
number of parametric pumping regions even though evanescent modes may exist
in such regions.

\end{document}